\documentclass[prl,aps,showpacs,twocolumn]{revtex4}
\usepackage{amsmath}
\usepackage{latexsym}
\usepackage{amssymb}
\usepackage{graphics,epstopdf}
\begin{document}
\title{Probing the Leggett-Garg Inequality for Oscillating  Neutral Kaons and Neutrinos}
\author{D. Gangopadhyay$^{1}$ (debashis@rkmvu.ac.in), D. Home$^{2}$ (dhome@bosemain.boseinst.ac.in) and A. Sinha Roy$^{1}$ (animesh@rkmvu.ac.in)}
\affiliation{$^1$Department of Physics, Ramakrishna Mission Vivekananda University, Belur Matth, Howrah-711202, India}
\affiliation{$^2$CAPSS, Department of Physics, Bose Institute, Sector-V, Salt Lake, Kolkata 700 091, India}
\begin{abstract}
The Leggett-Garg inequality (LGI) as a temporal analogue of Bell's inequality, derived using the notion of realism, is applied in a hitherto unexplored context involving the weak interaction induced two-state oscillations of decaying neutral kaons and neutrinos. The maximum violation of LGI obtained from the quantum mechanical results is significantly higher for the oscillating neutrinos compared to that for the kaons. Interestingly, the effect of CP non-invariance for the kaon oscillation is to enhance this violation while, for neutrinos, it is sensitive to the value of the mixing parameter.
\end{abstract}
\maketitle
\textit{Introduction}-Nonclassical features of the quantum world and their deeper implications have been explored since the inception of quantum mechanics (QM). A novel slant on this line of study was provided by Bell's inequality (BI)\cite{bell} whose distinctiveness stems from the feature that it is a testable algebraic consequence of a very basic notion known as local realism. This assumes that observables pertaining to any object, even when not measured, have definite values (the notion of realism), and that results of individual measurements of these properties remain unaffected by spatially distant events (the locality condition). BI essentially sets a bound on a certain combination of correlation functions corresponding to outcomes of measurements on two spatially separated systems. For suitable relative orientations of these measurements, BI is violated by the relevant QM results for appropriate states of the entangled systems. Extensive experimental investigations \cite{aspect} over the past three decades have succeeded in closing all possible loopholes, thereby providing convincing empirical repudiation of BI, consistent with the QM predictions. The upshot of these studies is the realization that any realist description of microphysicone of the earliest examples of CP violation in natureal phenomenon has to be intrinsically nonlocal.

A stimulating twist to the above line of study was provided by the Leggett-Garg inequality (LGI) \cite{leggett, leggett1} which is a temporal analogue of BI in terms of time-separated correlation functions corresponding to successive measurement outcomes for a system whose state evolves in time. Notion of realism is invoked in deriving LGI by assuming that a system, during its time evolution, is at any given time in a definite one of the available states. Instead of the locality condition, the notion of noninvasive measurability (NIM) is used. This means assuming that it is possible, in principle, to determine which of the states the system is in, without affecting the state itself or the system's subsequent dynamics. Leggett \cite{leggett1} has argued justifying why NIM is to be considered a ``natural corollary'' of the notion of realism. QM violation of LGI in suitable examples would, therefore, signify repudiation of the notion of realism that includes the assumption of NIM. Thus, while furnishing a signature of distinctly quantum behaviour, LGI can be regarded as complementing BI in providing valuable insight into the nature of physical reality that is entailed by nonclassicality of quantum systems. Hence it has been of considerable interest to investigate the extent to which LGI is violated by QM for various types of systems. The original motivation \cite{leggett, leggett1} that led to LGI was to use it for probing the limits of quantum mechanics in the macroscopic regime; e.g., in the context of suitable experiments involving the rf-SQUID device \cite{van}. In recent years, study of QM violation of LGI has been carried out for a variety of solid-state qubit systems \cite{ruskov} and optical systems \cite{gossin}. Empirical violations of LGI have been shown using `weak measurements' \cite{laloy}, employing liquid-state nuclear magnetic resonance in chloroform \cite{souza}, as well as for single spins in a diamond defect center \cite{waldherr} and for nuclear spins precessing in an external magnetic field \cite{athalye}.

Against this backdrop, the present paper initiates studies along an earlier unexplored direction. Neutral kaons and neutrinos are particularly interesting systems in elementary particle physics which provide remarkable examples of oscillations in time over different states - an inherent property of these systems which is governed by the fundamental electro-weak interaction. While the phenomenon of neutrino oscillation \cite{fuku} has played a vital role in establishing non-zero rest mass of neutrinos, the significance of kaon oscillation \cite{gell} is that it involves eigenstates of the effective weak interaction Hamiltonian which exhibit decay embodying CP non-invariance - the earliest detected example of CP violation in nature. For these systems, it is, therefore, of prime interest to investigate not only the extent to which QM results can violate LGI, but also the nature of dependence of QM violation of LGI on certain relevant key parameters, such as the magnitude of CP violation for kaon oscillation and the mixing angle for neutrino oscillation. Here it is relevant to note that an incompatibility between BI and the relevant QM results for entangled pairs of neutral kaons has been demonstrated using Wigner's version of Bell's inequality \cite{bramon}, while there were also earlier attempts \cite{six} towards demonstrating  QM violation of local realism using neutral kaons. In view of such studies, it is thus of added relevance to examine the QM violation of a temporal analogue of BI for oscillating neutral kaons.  

\textit{The Leggett-Garg inequality} - We begin by setting up the relevant form of LGI that will be used in this paper. For this, we focus on a two-state system whose temporal evolution consists of oscillations between the states, say, 1 and 2. Let Q(t) be an observable quantity such that, whenever measured, it is found to take a value $+1(-1)$ depending on whether the system is in the state 1(2). Next, consider a collection of runs starting from identical initial conditions such that on the first series of runs Q is measured at times $t_{1}$ and $t_{2}$, on the second at $t_{2}$ and $t_{3}$, on the third at $t_{3}$ and $t_{4}$, and on the fourth at $t_{1}$ and $t_{4}$ (here $t_{1}<t_{2}<t_{3}<t_{4}$). From such measurements, it is straightforward to determine the temporal correlations $C_{ij} \equiv \langle Q(t_{i})Q(t_{j})\rangle$. Now, as argued by Leggett and Garg \cite{leggett, leggett1}, it is possible to adapt in this context, the standard argument leading to a Bell-type inequality with the times $t_{i}$ playing the role of apparatus settings. One can then use the following consequence of the assumptions of realism and NIM that were mentioned earlier. For any set of runs corresponding to the same initial state, any individual $Q(t_i)$ has the same definite value, irrespective of the pair \textbf{$Q(t_{i}) Q(t_{j})$} in which it occurs; i.e., the value of $Q(t_i)$ in any pair does not depend on whether any prior or subsequent measurement has been made on the system. Consequently, the combination $[Q(t_{1})Q(t_{2}) + Q(t_{2})Q(t_{3}) + Q(t_{3})Q(t_{4}) - Q(t_{1})Q(t_{4})]$ is always +2 or $-2$. If all the individual product terms in this expression are replaced by their averages over the entire ensemble of such sets of runs, the following form of LGI is then obtained
\begin{equation}
\label{LG1}
C \equiv C_{12} + C_{23} + C_{34} - C_{14} \leq 2
\end{equation}
The above is, thus, an inequality imposing realist constraints on the time-separated joint probabilities pertaining to oscillations in any two-state system. We now analyse its incompatibility with QM in the specific cases of kaon and  neutrino oscillations respectively.

\textit{Oscillating neutral kaons} - Neutral kaons $K^0,\bar{K^0}$ are pseudo scalar mesons, each of which is the antiparticle of the other and is distinguished by the strangeness quantum number S = +1 (-1) for $K^0$ $(\bar{K^0})$. Evolving under the effective weak interaction $CP$ violating Hamiltonian $H=M-i\Gamma/2$ where M and $\Gamma$ are the Hermitian mass and decay matrices respectively, $K^0$ and $\bar{K^0}$ decay, while giving rise to $K^0$,$\bar{K^0}$ oscillations. Eigenstates of $H$ are respectively the long-lived ($\tau_{L}\sim 10^{-7}s$) and short-lived($\tau_{S}\sim 10^{-10}s$)states $|K_{L,S}\rangle$ with eigenvalues $\lambda_{L,S}=m_{L,S}-i\gamma_{L,S}/2$ corresponding to masses $m_{L,S}$ and characteristic decay widths $\gamma_{L,S}(=\hbar/\tau_{L,S})$ where $|K_{L,S}(t)\rangle=exp(-i\lambda_{L,S}t)|K_{L,S}\rangle$. In terms of mutually orthogonal strangeness eigenstates $K^{0},\bar{K^{0}}$ one has
\begin{eqnarray}
\label{LG2}
\vert{K_{L,S}}\rangle &=& \frac{1+\epsilon}{\sqrt{2(1+\vert{\epsilon}\vert ^{2})}}\vert{K^0}\rangle \pm \frac{1-\epsilon}{\sqrt{2(1+\vert{\epsilon}\vert ^{2})}}\vert{\bar{K^0}}\rangle
\end{eqnarray}
\noindent where $\epsilon$ is the complex CP violating parameter and $\langle K_L|K_S\rangle = 2Re(\epsilon)/(1+|\epsilon|^{2})$. Using Eq.(\ref{LG2}), and the time evolution of $|K_{L,S}\rangle$, one can calculate the time evolution of an initial pure $K^0$ beam. The probabilities of finding $K^0$ and $\bar{K^0}$ states after time t are respectively
\begin{eqnarray}
\label{LG3}
P_{K^0}(t)=\frac{e^{-\gamma_Lt}+e^{-\gamma_St} +2e^{-\gamma t}\cos(\Delta mt)}{4}
\end{eqnarray}
\begin{eqnarray}
\label{LG3a}
P_{\bar{K^0}}(t)=\frac{|({1-\epsilon})|^{2}}{|({1+\epsilon})|^{2}}\left[\frac{e^{-\gamma_Lt}+e^{-\gamma_St} -2e^{-\gamma t}\cos(\Delta mt)}{4}\right]
\end{eqnarray}
\noindent where $\Delta m = m_{L}-m_{S}$ and $\gamma = (\gamma_{S}+\gamma_{L})/2$.
The joint probability $P_{K^0,\bar{K^0}}(t_1,t_2)$ of finding the states $|K^0\rangle,\bar{|K^0\rangle}$ at the respective times $t_1$ and $t_2$ is then
\begin{eqnarray}
\label{LG4}
P_{K^0,\bar{K^0}}(t_1,t_2)=&\frac{1}{4}\biggl[e^{-\gamma_Lt_1}+e^{-\gamma_St_1} +2e^{-\gamma t_1}\cos(\Delta mt_1)\biggr]\nonumber\\
&\times \frac{|({1-\epsilon})|^{2}}{4|({1+\epsilon})|^{2}}\biggl[e^{-\gamma_L(t_2-t_1)}+e^{-\gamma_S(t_2-t_1)}\nonumber\\
& -2e^{-\gamma (t_2-t_1)}\cos\{\Delta m(t_2-t_1)\}\biggr]
\end{eqnarray} 
Similarly, one can calculate the other joint probabilities $ P_{K^0,K^0}(t_1,t_2), P_{\bar{K^0},K^0}(t_1,t_2)$ and $P_{\bar{K^0},\bar{K^0}}(t_1,t_2) $. Then the time correlation function  $C_{12}(\equiv\langle Q(t_1)Q(t_2)\rangle)$ can be evaluated by using expressions like Eq. (\ref{LG4}) to obtain 
\begin{eqnarray}
\label{LG6}
C_{12} =&\biggl[\frac{(1+|\epsilon |^{2})(e^{-\gamma_Lt_1}+e^{-\gamma_St_1})}{2}e^{-\gamma(t_2-t_1)}\cos\{\Delta m(t_2-t_1)\}\nonumber\\
&+Re(\epsilon)e^{-\gamma t_1}\{e^{-\gamma _L(t_2-t_1)}+e^{-\gamma _S(t_2-t_1)}\}\cos(\Delta mt_1)\biggr]\nonumber\\
&\times \biggl[\frac{(1+|\epsilon |^{2})(e^{-\gamma_Lt_1}+e^{-\gamma_St_1})}{4}\{e^{-\gamma _L(t_2-t_1)}+e^{-\gamma _S(t_2-t_1)}\}\nonumber\\
&+2Re(\epsilon)e^{-\gamma t_2}\cos(\Delta mt_1)\cos\{\Delta m(t_2-t_1)\}\biggr]^{-1}
\end{eqnarray} 
An interesting point to be noted here is that, in the presence of CP violation $(\epsilon\neq 0)$, $C_{12}$ displays dependence on both the temporal separation $(t_2 - t_1)$ and on $t_1$. If $\epsilon=0$, the dependence on $t_1$ disappears, as can be checked from Eq.(\ref{LG6}). The other temporal correlation functions $C_{23}, C_{34}$ and $C_{14}$ can be calculated in the same way and they also show a similar feature. Here we may note that under CPT invariance, CP violation implies time reversal non-invariance. This explains why in the presence of CP violation, the $C_{ij}'s$ are not merely functions of the temporal separation but also depend on the initial instant of the pertinent time interval.
 
Next, using equations like Eq.(\ref{LG6}), one can evaluate C as defined in Eq.(\ref{LG1}) in order to study the QM violation of LGI for kaon oscillations. By varying the choices of the time intervals, it is found that the maximum value of C is attained essentially when the temporal intervals are chosen to be the same, i.e., when $t_{4} - t_{3} = t_{3} - t_{2} = t_{2} - t_{1} = \Delta t$. The detailed expression for C under this condition and in the presence of CP violation is given in the \textit{supplementary material}\cite{supple} where it can be seen explicitly that C depends on both $\Delta t$ and $t_{1}$. The experimentally determined values of $\Delta{m},\tau _s,\tau _L,|\epsilon|$ and $Re(\epsilon)$ occurring in the expression for C are $\tau _s=0.8958\times 10^{-10}\text{sec}$, $\tau _L=0.5084\times 10^{-7}\text{sec}$, $Re(\epsilon)=1.596\times 10^{-3}$ \cite{ambros}, while $\Delta m=3.843\times 10^{-12}\text{MeV}$, $|\epsilon| = 2.232\times 10^{-3} $ \cite{yao}. We consider various choices of $t_{1}$ and $\Delta t$ within an appropriate time scale ($t_{1}/\tau_{S}\leq 10, \Delta t/\tau_{S}\lesssim 10$) over which the decaying kaons show appreciable oscillation. It is then found that the QM calculated \textit{maximum value} of C in the oscillation time range is 2.36463 when $\Delta t = 0.7889\times\tau_{S}$ and $ t_{1} = 5.3\times\tau_{S}$. Note that the maximum QM violation of LGI in this case is significantly smaller than the upper bound \cite{cirel} of the QM violation of the Bell-type inequality given by $2\sqrt{2}=2.82843$. For a given value of $t_{1} = 5.3\times\tau_{S}$, the variation of QM calculated quantity C with $\Delta t/\tau_{S}$ is shown as the unbroken curve in Fig.1(a). Here we observe that, if, for example, the difference between the decay parameters $(\tau_{L}-\tau_{S})$ had a lower value, the maximum QM value of C would have increased.\\
\begin{figure}
\resizebox{8.5cm}{5.5cm}{\includegraphics{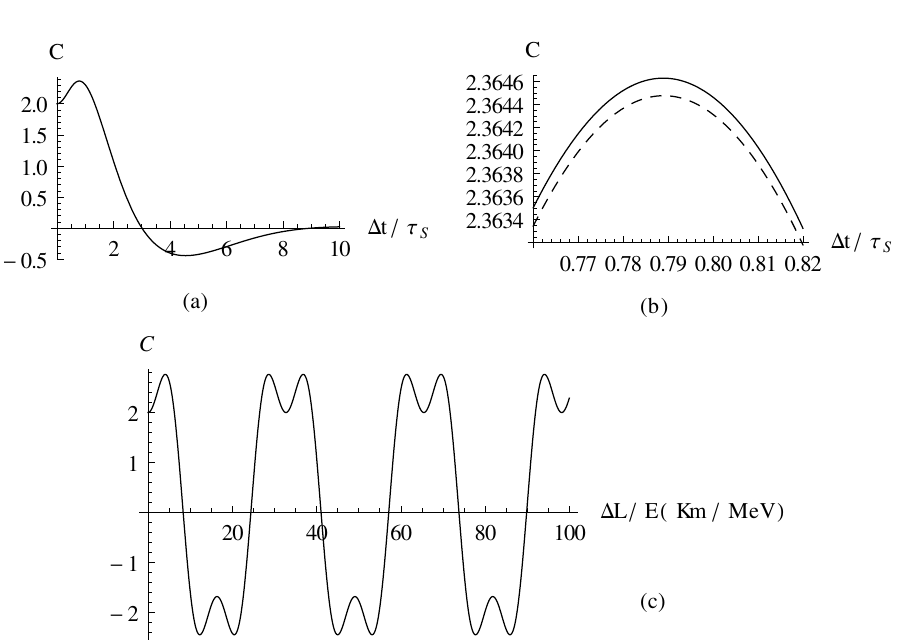}}
\caption{In Fig.1(a) pertaining to kaon oscillation in the presence of CP violation, the overall behaviour of the QM calculated value of $C$ as a function of ($\Delta t/\tau_{S}$) is shown within the oscillation time range, where we take $t_{1}/\tau_{S}=5.3$. In Fig.1(b), we focus around the region where the QM value of $C$ is maximum as ($\Delta t/\tau_{S}$) is varied. Comparing the unbroken and broken curves for the cases with and without CP violation, one can see the enhancement of the QM violation of LGI due to CP violation. In Fig.1(c), pertaining to neutrino oscillation, we show the overall behaviour of the QM value of C as the parameter $\Delta L/E$ is varied for the KamLAND experimental setup.}
\end{figure}
If we ignore the effect of CP violation and put $\epsilon = 0$, C depends only on $\Delta t$ as can be seen from the detailed expression for C given in the \textit{supplementary material}\cite{supple}. By varying $\Delta t$, it is then found that the \textit{maximum QM value} of C in the oscillation time range is $2.36448$ for $\Delta t = 0.7890\times \tau_{S}$, whereas, as mentioned earlier, the maximum value of C in the presence of CP violation is $2.36463$. Thus, while the overall behaviour of the QM calculated quantity C as $\Delta t/\tau_{S}$ is varied is the same with and without CP violation, a notable feature is that the QM violation of LGI is enhanced by an amount 0.00015 in the presence of CP violation (see Fig. 1(b)). We have also checked that if $|\epsilon |$ was, for example, larger, say, $2.23\times 10^{-2}$, the maximum QM value of C would have increased to 2.36667. Conceptually, this means that the non-invariance of $CP$ symmetry in weak interaction results in an enhanced non-classicality of the kaon oscillation as shown by an increased QM violation of LGI.

\textit{Oscillating neutrinos} - It has been well established by a number of experiments that during propagation, neutrinos undergo oscillations among three flavor eigenstates $\nu_{e},\nu_{\mu},\nu_{\tau}$. Here our treatment of LGI using the neutrino oscillation phenomenon is analysed in the context of the KamLAND experimental setup where the effect of the mixing angle $\theta_{13}$ can be considered negligibly small \cite{abe}. Then for neutrinos, one can essentially consider two-flavor oscillation involving transitions between $\nu_{e}$ and $\nu_{\mu}$. Here the mass eigenstates are $\nu_{1}$ and $\nu_{2}$ with energy eigenvalues $E_{1,2}$ respectively, where $\nu_1=\cos{\theta} \nu_{\mu} -\sin{\theta} \nu_{e}$, $\nu_2=\sin{\theta} \nu_{\mu} +\cos{\theta} \nu_{e}$ and $\nu_{1,2}(t)=\nu_{1,2}exp(-iE_{1,2}t/\hbar)$. Now, given an initial electron neutrino beam, after time t, the probability of obtaining $\nu_{\mu}$ and $\nu_{e}$ are respectively \cite{griffith}
\begin{eqnarray}
\label{LG7}
P_{\nu_e\rightarrow\nu_\mu}(t)= \biggl[\sin(2\theta)\sin\biggl(\frac{\Delta m^{2}}{4\hbar E}c^{4}t\biggr)\biggr]^{2}
\end{eqnarray}
\begin{eqnarray}
\label{LG8}
P_{\nu_e\rightarrow\nu_e}(t)=1- \biggl[\sin(2\theta)\sin\biggl(\frac{\Delta m^{2}}{4\hbar E}c^{4}t\biggr)\biggr]^{2}
\end{eqnarray}
\noindent where $\theta$ is the mixing angle, $\Delta m^{2}=m_{2}^{2}-m_{1}^{2}$ where $m_{1}$ and $m_{2}$ are masses corresponding to the states $\nu_{1}$ and $\nu_{2}$ respectively and E is the mean energy of neutrino mass eigenstates. Next, one can calculate the four joint probabilities (similar to those defined for the kaon oscillation) $P_{\nu_e,\nu_\mu}(t_1,t_2),P_{\nu_{e},\nu_{e}}(t_1,t_2), P_{\nu_{\mu},\nu_{e}}(t_1,t_2)$ and $P_{\nu_{\mu},\nu_{\mu}}(t_1,t_2)$. The time correlation function $C_{12}$ in this case is given by
\begin{eqnarray}
\label{LG10}
C_{12}=\left[1-2 \biggl\{\sin(2\theta)\sin\biggl(\frac{\Delta m^{2}}{4\hbar E}c^{4}(t_2-t_1)\biggr)\biggr\}^{2}\right]
\end{eqnarray}
Similarly, the functions $C_{23}, C_{34}$ and $C_{14}$ can be calculated. Here the temporal correlation functions depend only on the temporal separation. Next, using equations like Eq.(\ref{LG10}), one can evaluate the quantity C as defined in Eq.(\ref{LG1}) in order to study the QM violation of LGI for the oscillating neutrinos. By varying the choices of the time intervals, similar to the case of kaon oscillation, it is found that the maximum QM value of C for the neutrino oscillation, too, occurs only when $t_{4} - t_{3} = t_{3} - t_{2} = t_{2} - t_{1} = \Delta t$. The quantity C under this condition becomes a function of $\Delta t/E$. Now, if $\Delta L$ be the distance traversed by the neutrinos during time interval $\Delta t$, then C is given by
\begin{eqnarray}
\label{LG11}
C=&&2-2\sin^{2}2\theta\biggl[3\sin^{2}\{\frac{\Delta m^{2}c^{4}}{4\hbar cE}\Delta L\}\nonumber\\
&&-\sin^{2}\{\frac{3\Delta m^{2}c^{4}}{4\hbar cE}\Delta L\}\biggr]
\end{eqnarray}
\noindent with $\Delta L=c\Delta t$. Since neutrinos are non-decaying, it is clear from Eq.(\ref{LG11}) that, unlike in the case for neutral kaons, the QM calculated quantity C continues to oscillate with time with maxima points occurring at various points of $(\Delta L/E)$ . The experimentally obtained values of $\Delta m^{2}c^{4},\tan^{2}\theta$ in the case of the KamLAND experimental setup (involving reactor neutrinos and geo neutrinos) are respectively $7.58\times 10^{-5} ev^{2}$ and $0.56$ \cite{abe}. For such data, the maximum QM value of C is 2.76 and this maximum value is repeated at various points of $(\Delta L/E)$; see Fig. 1(c). Further, we have checked by varying $\Delta m^{2}$ over a range of values that there is no significant change from the maximum QM value of $C$. On the other hand, the maximum QM value of C depends sensitively on the value of the mixing angle $\theta$. If $\theta$ had a higher value than the experimentally determined value, the maximum QM value of C would have increased, reaching the upper bound $2\sqrt{2}$ for $\theta=\pi/4$, and then would have decreased if the value of $\theta$ was still higher.

\textit{Concluding remarks} - Comparing the magnitudes of the QM violation of LGI in the cases of the kaon and two-flavour neutrino oscillation, it is found that while in the former case, the ratio by which the realist upper bound given by LGI is violated by the relevant QM results is $18\%$, this ratio is $38\%$ for the neutrino oscillation, thereby significantly higher than the corresponding value for the kaon oscillation, but smaller than the  upper bound of this ratio (i.e.,$41\%$) by which LGI can be violated by QM. On the other hand, for the kaon oscillation, the effect of CP violation is to enhance the QM violation of LGI by $0.008\%$ - albeit small, but not a negligible effect that is conceptually interesting.

Possible implications of the above results call for careful reflection. Besides, it could be worthwhile to make a detailed quantitative comparison of the results obtained in this paper with the QM incompatibility with local realism analysed for neutral kaons \cite{bramon,six}. Finally, we should mention that the phenomenon of neutrino oscillation recently studied in the context of the Daya Bay reactor neutrino experimental setup \cite{an} involves a non-negligible value of the mixing angle $\theta_{13}$, thereby entailing three-flavour neutrino oscillation. It could, therefore, be interesting to examine whether the QM incompatibility with the notion of realism underpinning LGI can be probed in such a context as well.

A.S.R. thanks UGC-CSIR for providing a Research Fellowship Sr.No. 2061151173 under which this work was done. D.H. thanks DST Project No. SR/S2/PU-16/2007 and Centre for Science, Kolkata for supporting his research.
 

\begin{thebibliography} {99}
\bibitem{bell} J. S. Bell, Physics  (Long Island City, N. Y.) \textbf{1}, 195 (1964).
\bibitem{aspect} For example, A. Aspect, P. Grangier and G. Roger, Phys. Rev. Lett. \text{49}, 91 (1982); \textit{ibid}. \textbf{49}, 1804 (1982); W. Tittel, J. Brendel, H. Zbinden and N. Gisin, Phys. Rev. Lett. \textbf{81}, 3563 (1998); G. Weihs, T. Jennewein, C. Simon, H. Weinfurter, and A. Zeilinger, Phys. Rev. Lett. \textbf{81}, 5039 (1998); M. A. Rowe, D. Kielpinski, V. Meyer, C. A. Sackett, W. M. Itano, C. Monroe, and D. J. Wineland, Nature (London) \textbf{409}, 791 (2001); D. Salart, A. Baas, J. A. W. van Houwelingen, N. Gisin, and H. Zbinden, Phys. Rev. Lett. \textbf{100}, 220404 (2008).
\bibitem{leggett} A. J. Leggett and A. Garg, Phys. Rev. Lett. \textbf{54}, 857 (1985).
\bibitem{leggett1} A. J. Leggett, J. Phys. Condens. Matter \textbf{14}, R415 (2002); Rep. Prog. Phys. \textbf{71}, 022001 (2008).
\bibitem{van} C. H. van der Wal \textit{et al.}, Science \textbf{290}, 773 (2000); J. R. Friedman \textit{et al.}, Nature \textbf{406}, 43 (2002).
\bibitem{ruskov} R. Ruskov, A. N. Korotkov and A. Mizel, Phys. Rev. Lett. \textbf{96}, 200404 (2006); A. N. Jordan, A. N. Korotkov and M. Buttiker, Phys. Rev. Lett. \textbf{97}, 026805(2006); N. S. Williams and A. N. Jordan, Phys. Rev. Lett. \textbf{100}, 026804 (2008); G. C. Knee \textit{et al.}, Nature Communications \textbf{3}, 606 (2012).
\bibitem{gossin} M. E. Gossin \textit{et al.}, Proc. Natl Acad. Sci. \textbf{108}, 1256 (2011).
\bibitem{laloy} A Palacios-Laloy \textit{et al.}, Nature Physics \textbf{6}, 442 (2010); J. Dressel \textit{et al.}, Phys. Rev. Lett. \textbf{106}, 040402 (2011). 
\bibitem{souza} A. M. Souza, I. S. Oliveira and R. S. Sarthour, New J. Phys. \textbf{13}, 053023 (2011).
\bibitem{waldherr} G. Waldherr \textit{et al.}, Phys. Rev. Lett. \textbf{107}, 090401 (2011).
\bibitem{athalye} V. Athalye, S. S. Roy and T. S. Mahesh, Phys. Rev. Lett. \textbf{107}, 130402 (2011).
\bibitem{fuku} Y. Fukuda \textit{et al.}, Phys. Rev. Lett. \textbf{81}, 1562 (1998).
\bibitem{gell} M. Gell-Mann and A. Pais, Phys. Rev. \textbf{97}, 1387 (1955).
\bibitem{bramon} A. Bramon and M. Nowakowski, Phys. Rev. Lett. \textbf{83}, 1 (1999).
\bibitem{six} J. Six, Phys. Rev. Lett. B \textbf{114}, 200 (1982); A. Datta and D. Home, Phys. Lett. A \textbf{119}, 3 (1986); Found. Phys. Lett. \textbf{4}, 165 (1991); P. A. Eberhard, Nucl. Phys. \textbf{B398}, 155 (1993); A. Di Domenico, Nucl. Phys.  \textbf{B450}, 293 (1995); F. Selleri, Phys. Rev. A \textbf{56}, 3493 (1997).
\bibitem{supple} See attached supplementary material.
\bibitem{ambros} F. Ambrosino \textit{et al.}, JHEP \textbf{12}, 011 (2006).
\bibitem{yao} W-M Yao \textit{et al.}, J. Phys. G: Nucl. Part. Phys. \textbf{33}, 1 (2006).
\bibitem{cirel} B. S. Cirel'son, Lett. Math. Phys. \textbf{4}, 93 (1980).
\bibitem{abe} S. Abe, \textit{et al.}, Phys. Rev. Lett. \textbf{100}, 221803 (2008).
\bibitem{griffith} David J. Griffiths, \textit{Introduction to Elementary Particles} ( Wiley-VCH,  Weinheim, 2008), Chap. 11, p. 392.
\bibitem{an} F. P. An, \textit{et al.}, Phys. Rev. Lett. \textbf{108}, 171803 (2012).
\end{thebibliography}
\end{document}